\documentclass[aps,prd,twocolumn,a4paper,superscriptaddress,floatfix]{revtex4-1}
\usepackage{graphicx,color,amsmath}
\begin{document}

\newcommand{\be}{\begin{equation}}
\newcommand{\ee}{\end{equation}}
\newcommand{\bq}{\begin{eqnarray}}
\newcommand{\eq}{\end{eqnarray}}
\newcommand{\bw}{\begin{widetext}}
\newcommand{\ew}{\end{widetext}}
\newcommand{\ba}{\begin{align}}
\newcommand{\ea}{\end{align}}
\newcommand{\bc}{\begin{center}}
\newcommand{\ec}{\end{center}}

\title{Scaling solutions of wiggly cosmic strings: II. Time-varying coarse-graining scale solutions}
\author{A. Almeida}
\email{Ana.Almeida@astro.up.pt}
\affiliation{Centro de Astrof\'{\i}sica da Universidade do Porto, Rua das Estrelas, 4150-762 Porto, Portugal}
\affiliation{Instituto de Astrof\'{\i}sica e Ci\^encias do Espa\c co, CAUP, Rua das Estrelas, 4150-762 Porto, Portugal}
\affiliation{Faculdade de Ci\^encias, Universidade do Porto, Rua do Campo Alegre 687, 4169-007 Porto, Portugal}
\author{C. J. A. P. Martins}
\email{Carlos.Martins@astro.up.pt}
\affiliation{Centro de Astrof\'{\i}sica da Universidade do Porto, Rua das Estrelas, 4150-762 Porto, Portugal}
\affiliation{Instituto de Astrof\'{\i}sica e Ci\^encias do Espa\c co, CAUP, Rua das Estrelas, 4150-762 Porto, Portugal}

\date{\today}

\begin{abstract}
We continue our exploration of the wiggly generalisation of the Velocity-Dependent One Scale Model for cosmic strings, through the study of its allowed asymptotic scaling solutions. We extend the work of a previous paper [Almeida $\&$ Martins, Phys. Rev. D 104 (2021) 043524] by considering the more comprehensive case of a time-varying coarse-graining scale for the string wiggles. The modeling of the evolution of the network therefore relies on three main mechanisms: Hubble expansion, energy transfer mechanisms (e.g., the production of loops and wiggles) and the choice of the scale at which wiggles are coarse-grained. We analyse the role of each of them on the overall behaviour of the network, and thus in the allowed scaling solutions. In Minkowski space, we find that linear scaling, previously observed in numerical simulations without expansion, is not possible with a changing averaging scale. For expanding universes, we find that the three broad classes of scaling solutions---with the wiggliness disappearing, reaching scaling, or growing---still exist but are differently impacted by the time evolution of the coarse-graining scale. Nambu-Goto type solutions (without wiggles) are unaffected, growing wiggliness solutions are trivially generalized, while for solutions where wiggliness reaches scaling the expansion rate for which the solution exists is decreased with respect to the one for a fixed coarse-graining scale. Finally, we also show that the inclusion of a time-varying coarse-graining scale allows, in principle, for additional scaling solutions which, although mathematically valid, are not physical. Overall, our mapping of the landscape of the allowed scaling solutions of the wiggly Velocity-Dependent One Scale Model paves the way for the detailed testing of the model, to be done by forthcoming high-resolution field theory and Nambu-Goto simulations.
\end{abstract}
\maketitle

%%%%%%%%%%%%%%%%%%%%%%%%%%%%%%%%%%%%%%%%%%%%%%%
\section{\label{sint}Introduction}

Topological defects are predicted to have been formed during phase transitions in the very early universe, as a consequence of the Kibble mechanism \cite{Kibble76}. Depending on the topology of the vacuum manifold one can have various classes of defects, with cosmic strings being the best motivated and cosmologically more interesting ones. The study of cosmic string network evolution is therefore an important part of understanding the physics of the early phases of the universe \cite{VSH}.

However, due to the complexity inherent to string networks, a detailed quantitative understanding of cosmic string evolution is difficult. The best approach to this problem relies on the interplay between Nambu-Goto or Abelian-Higgs numerical simulations \cite{BB,AS,ABELIAN,FRAC,RSB,VVO,Stuckey,Blanco,Hiramatsu,Correia1,Correia2,Correia3} and analytic modelling. Analytic models are rigorously derived from first principles, using the string microphysics so as to derive equations that govern the evolution of the network expressed in terms of the relevant macroscopic quantities. The loss of information that comes with this procedure is encapsulated in the form of phenomenological parameters which have to be calibrated with simulation data. 

The most successful model of cosmic string evolution to date is the so-called Velocity-Dependent One Scale (VOS) Model \cite{MS2,MS3,MS4,Book}, which has been extensively and successfully tested against numerical simulations of cosmic strings \cite{ABELIAN,FRAC,Correia1,Correia3}; analogous models also exist for other topological defects \cite{Book}. The VOS is able to provide an accurate depiction of the large-scale behavior for the simplest cosmic string networks. Nevertheless, efforts to extend it for describing cosmologically more realistic networks of strings, whose string worldsheets are expected to have additional degrees of freedom, are more recent. This limitation is evident in the inability of the first-generation VOS to explicitly take into account the behavior of the small-scale structure that is known to build up in realistic string networks \cite{ACK,FRAC,POLR}. Numerical simulations have demonstrated the presence of non-negligible amounts of short-wavelength propagation modes (known as wiggles) at scales significantly below the correlation length. Because small-scale structure is a byproduct of the energy loss phenomena of the network, an accurate picture of cosmic string evolution can only be achieved if these small-scale dynamical processes are also modelled. This motivated a wiggly generalisation of the VOS \cite{PAP1,PAP2}, which explicitly describes the evolution of small-scale structure in the network, while also retaining the ability of the VOS to capture the large-scale properties of the network.

In a previous paper \cite{almeida} (henceforth denoted Paper I) we improved the physical interpretation of the wiggly model through an exploration of the mathematical landscape of asymptotic scaling solutions. Specifically, we focused on the case of a constant coarse-graining scale for the small-scale structures, and identified three classes of network scaling solutions, which describe physically different behaviours of the small-scale structure of the network. In addition to the Nambu-Goto solution (without wiggles), to which the wiggly model reduces in the appropriate limit, there are also solutions where the wiggliness of the network grows, and, under more specific conditions, also solutions where the wiggliness can itself reach scaling. Other things being equal, which of the three regimes occurs depends primarily on the expansion rate. One consequence of this is that the full scaling of the network, including the wiggliness, is more likely in the matter era than in the radiation epoch, which is agreement with numerical studies \cite{AS,BB,FRAC}. On the other hand, in Minkowski space, linear scaling is possible in the model, again in agreement with Minkowski space numerical simulations \cite{FRAC,Sakellariadou}.

The present work continues the study of the asymptotic scaling solutions of the wiggly model, now extending this analysis to cases where the coarse-graining scale is allowed to vary. Specifically, we consider averaging scales which vary as a power law of physical time. This choice is done in part for the sake of mathematical tractability, but also because it allows us to consider what is arguably the best physically motivated choice for such a non-constant scale: that of the correlation length, which in most circumstances is itself expected to vary as a power law of time. This allows us to address, in a wider parameter space, a question which was already the focus of Paper I: whether small-scale structure reaches scaling, and under what physical conditions this can happen. Additionally, this also enables us to further clarify the role of each physical mechanism on the evolution of the network. In this regard, Paper I focused on the role of the cosmological expansion rate (also including the particular case of Minkowski space) and of the network's energy transfer mechanisms (e.g., the production of loops and wiggles). In the present work the role of the time-varying averaging scale is also included in this analysis.

This paper is laid out as follows. We start with a brief review of the mathematical formalism underlying the wiggly extension of the VOS model in Section \ref{sevo}. In Sections \ref{flat} and \ref{expand}, we present the scaling solutions of the wiggly model in Minkowski and power-law expanding universes, respectively, and discuss their physical interpretation. This structure mirrors the one of Paper I, which will hopefully facilitate the comparison of the results of the two works---in other words, many (though not all) of the solutions to be discussed in what follows are extensions of solutions already presented in Paper I. Some of these solutions were also briefly reported in a recent conference proceedings \cite{Proceedings}. Lastly, our findings are summarised in Section \ref{concl}.

%%%%%%%%%%%%%%%%%%%%%%%%%%%%%%%%%%%%%%%%%%%%%%%%%%%%%%
\section{\label{sevo}The VOS model for Wiggly cosmic string Evolution}

In this section we provide a short introduction to the physical assumptions and mathematical formalism of the VOS model and its wiggly extension. This is a review of previous work in the literature, and in particular it is a shorter version of the discussion in Section II of Paper I, but it is presented here in order to make the present work reasonably self-contained, in particular by defining all the relevant variables.

The canonical VOS framework provides a quantitative description of the evolution of a string network in terms of two macroscopic quantities: a root-mean squared velocity $v$ and a characteristic length scale $L$ which is identified as the string \textit{correlation length} $\xi$ and the string \textit{curvature radius} $R$. The VOS retains the one-scale assumption of Kibble's original model \cite{KIB,BMOD}, but with the inclusion of a mean velocity as an additional dynamical variable, one is now able to make quantitative predictions in various cosmological epochs.

One starts by defining the total energy of the network $E$ and the root-mean-square (RMS) velocity $v$
\bq
E&=&\mu_0 a(\tau)\int\epsilon d\sigma \\
v^2&=&\frac{\int{\dot{\bf x}}^2\epsilon d\sigma}{\int\epsilon d\sigma}\,,
\label{eee}
\eq
where $\mu_0$ is the string mass per unit length. String networks comprise long (or infinite) strings and small closed loops. The following will concern long strings. On large scales, these can be treated as a Brownian random walk \cite{FRAC}, to which we can assign a characteristic length scale (or inter-string separation) $L$. One can then express the energy $E= \rho V$ in terms of a length scale
\be
\rho \equiv \frac{\mu_0}{L^2}\,.
\ee
The averaging procedure comes with a cost: the introduction of phenomenological parameters that account for the small-scale physics of the network. In particular, the fraction of energy lost into the production of loops is encoded in a \textit{loop chopping efficiency} parameter $c$, defined as
\begin{equation}
\left(\frac{d\rho}{dt}\right)_{\rm to\ loops}= c v\frac{\rho}{L}
\, . \label{rtl}
\end{equation}
In addition, the presence of small-scale wiggles on the strings motivates the inclusion of the momentum (or curvature) parameter $k(v)$, for which detailed descriptions can be found in \cite{MS3,Correia1}. The VOS evolution equations can then be shown to have the following form
\bq
2\frac{dL}{dt}&=&2HL(1 + {v^2})+c v \\ \label{evl0}
\frac{dv}{dt}&=&\left(1-{v^2}\right)\left[\frac{k(v)}{L}-2Hv\right]\,, \label{evv0}
\eq
where $H\equiv\dot{a}/a$ is the Hubble parameter.

The one-scale approximation underlying the original VOS framework, $L=\xi=R$, implies that the model is unable to accurately capture the dynamics at length scales below the characteristic length. This motivates the wiggly VOS extension, whose thorough mathematical derivation can be found in \cite{PAP1,PAP2}. Its starting point is noting that wiggly strings have an energy density in the locally preferred string rest frame (denoted $U$) and a local string tension (denoted $T$) which are not identical nor constants, as would be the case in the Nambu-Goto case. Specifically, one can define them to depend on a dimensionless parameter $w$ that ranges from $0$ to $1$ (unity being the value of a Nambu-Goto string), such that 
\bq
T&=&w\mu_0\\ U&=&\frac{\mu_0}{w}\, ,\label{wigt}
\eq
and therefore $T/U=w^2$.

This motivates the redefinition of the total energy
\begin{equation}
E=a\int\epsilon Ud\sigma=\mu_0 a\int\frac{\epsilon}{w}d\sigma\, .\label{enertot}
\end{equation}
which is due to two main contributions: one from the bare strings
\begin{equation}
E_0=\mu_0 a\int\epsilon d\sigma\,, \label{enerst}
\end{equation}
with the rest lying in the small-scale wiggles. Naturally, one can assign characteristic length scales to each energy contribution. The string correlation length is defined with respect to the bare string density
\begin{equation}
\rho_0\equiv\frac{\mu_0}{\xi^2} \,, \label{wig_b2}
\end{equation}
and is a measure of the characteristic length of a Brownian network; the string wiggle density is denoted $\rho_w$. While the correlation length is still physically meaningful, the characteristic length scale $L$ only serves as a proxy for the total energy in the network,
\begin{equation}
\rho\equiv\frac{\mu_0}{L^2} \,, \label{wig_b2newL}
\end{equation}
which is trivially the sum of the bare and wiggle densities. A quantitative description of small-scale structure evolution is accomplished by introducing an additional quantity in the model, a renormalized string mass per unit length $\mu$, which is a measure of the energy due to the wiggliness of the network. It is natural to define it as a ratio between the total energy of the network and the energy in the bare string segments 
\begin{equation}
\mu\equiv\frac{E}{E_0}=\langle w\rangle^{-1}\,, \label{vwwv}
\end{equation}
or, equivalently, a ratio between length scales 
\begin{equation}
\xi^2=\mu L^2 \,.\label{ximuL}
\end{equation}
Evidently, $\mu$ is expected to take values greater than the unity, with $\mu=1$ corresponding to the Nambu-Goto case. Moreover, from Eq. (\ref{ximuL}) it becomes clear that we depart from the one-scale assumption of the VOS, as we now have two distinct length scales which will have distinct evolution. Consequently, our averaged model for wiggly cosmic string evolution will entail three independent differential equations, as opposed to two. In the last equality of Eq. (\ref{vwwv}), the $<\cdots>$ denotes an average over the string network. For a generic quantity $Q$, this is defined as
\begin{equation}
\langle Q\rangle=\frac{\int Q U \epsilon d \sigma}{\int U \epsilon d \sigma}=\frac{\int Q \frac{\epsilon}{w} d \sigma}{\int \frac{\epsilon}{w} d \sigma}\,,
\end{equation}
which attributes more weight to string segments with greater mass currents.

We also need additional phenomenological terms that model the energy transfers within the network, as some of them contribute to the generation of small-scale structure, while others are instrumental in its loss. Long string intercommutings increase the number of kinks on the string network, thus transferring energy to the wiggles. This process can be modelled as
\begin{equation}
\left(\frac{1}{\rho_0}\frac{d\rho_0}{dt}\right)_{\rm wig}= -cs(\mu)\frac{v}{\xi}\, , \label{wig_b6}
\end{equation}
such that $s$ vanishes in the Nambu-Goto limit and should also account for kink decay by gravitational radiation. Although kink formation occurs independently of loop production, intercommutings can also lead to the formation of loops. In fact, Nambu-Goto numerical simulations \cite{BB,AS,FRAC} suggest that small-scale structure might stimulate loop production. Phenomenologically, this translates into a function $f_0$ which explicitly depends on $\mu$. By analogy with Eq. (\ref{rtl}), we define the fraction of the bare energy density lost into loops per unit time as
\begin{equation}
\left(\frac{1}{\rho_0}\frac{d\rho_0}{dt}\right)_{\rm loop}=-c f_0(\mu) \frac{v}{\xi}\,; \label{wig_b4}
\end{equation}
in the Nambu-Goto limit $\mu \to 1$ $f_0$ should approach unity, such that Eq. (\ref{rtl}) is retrieved. Finally, one must also take into account that a fraction of the total energy lost into loops comes from the small-scale wiggles 
\begin{equation}
\left(\frac{1}{\rho_w}\frac{d\rho_w}{dt}\right)_{\rm loop}=-c f_1(\mu) \frac{v}{\xi}\,.
\label{wig_b5b}
\end{equation}
All in all, the total energy lost into loop the production is given by
\begin{equation}
\left(\frac{1}{\rho}\frac{d\rho}{dt}\right)_{\rm loop}=\left(\frac{1}{\rho}\frac{d\rho_0}{dt}\right)_{\rm loop}+\left(\frac{1}{\rho}\frac{d\rho_w}{dt}\right)_{\rm loop}\equiv -c f(\mu) \frac{v}{\xi}\,, \label{wig_b5a}
\end{equation}
where, for the sake of simplicity, in the last equality we have defined an overall loss parameter, $f$, which also has a dependence on $\mu$ so as to account for the fact that loop production is favoured in regions of the network containing more small-scale structure than average. As in Paper I, and following the earlier discussion in \cite{PAP2}, we assume that the energy loss parameters take the form 
\be\label{enloss}
\begin{split}
f_0(\mu)&=1\\
f(\mu)&=1+\eta\left(1-\frac{1}{\sqrt{\mu}}\right)\\ s(\mu)&=D\left(1-\frac{1}{\mu^2}\right)\,,
\end{split}
\ee
with $\eta$ and $D$ being new phenomenological parameters that can be understood as probabilities for small-scale structure loss and gain, respectively.

Lastly, it should be noted that $\mu$ has an explicit dependence on time, but also on the coarse-graining scale, denoted $\ell$; in other words, $\mu = \mu(\ell,t)$ \cite{FRAC}. The analysis of Paper I was limited to the case $\ell=const$; in what follows, we discuss the general case where $\ell$ can itself be time-dependent.

This coarse-graining scale (which can be, approximately but not exactly, envisaged as a renormalization scale in the particle physics sense) should be understood as a scale below the correlation length that is also large enough so that spatial variations in the energy density can be neglected. In other words, this provides a mesoscopic scale in the analytic model, which for that reason can no longer be a purely one-scale model. Physically, a change in this scale simply modifies the way the network energy is distributed between the bare string and the small-scale wiggles, while the total energy of the network is unaffected by this division. In other words, with a varying coarse-graining scale, scaling solutions are expected to be scale-dependent in the sense that the previously defined $v$ and $\xi$ (or equivalently $\mu$) should all be dependent on the coarse-graining scale. On the other hand one does not expect such a scale dependence for the characteristic length scale $L$, which is simply a measure of the total energy of the network and therefore should be independent of $\ell$. Indeed, solutions exhibiting such a dependence in $L$ should be considered non-physical.

Changing the coarse-graining scale is equivalent to redefining what small-scale structure is, and thus will not only affect the value of $\mu$ but also those of $E_{0}$ (equivalently, $\rho_0$ or $\xi$) and of $v$. This is accounted for by introducing the following scale-drift terms
\begin{equation}
\frac{1}{\mu}\frac{\partial\mu}{\partial\ell}\frac{d\ell}{dt}\sim\frac{d_{m}-1}{\ell}\frac{d\ell}{dt}\label{dm}
\end{equation}
\begin{equation}
\frac{\partial v^{2}}{\partial\ell}\frac{d\ell}{dt}=\frac{1-v^{2}}{1+\left\langle w^{2}\right\rangle}\frac{\partial\left\langle w^{2}\right\rangle}{\partial\ell}\frac{d\ell}{dt},\label{v_dm}
\end{equation}
where $d_{m}\left(\ell\right)$ is the multifractal dimension of a string segment at scale $\ell$ \cite{TAKAYASU}. Note that Eq.~(\ref{dm}) is a mere geometric identity, whereas Eq.~(\ref{v_dm}) ensures energy conservation at all scales.

With these definitions, and with the further assumption of uniform wiggliness (in other words, that $w$ varies only in time), one can obtain the evolution equations for the wiggly extension of the VOS model, which have the following form
\begin{widetext}
\begin{equation}
2\frac{dL}{dt}=HL\left[3+v^2-\frac{(1-v^2)}{\mu^2}\right]+\frac{cfv}{\sqrt{\mu}} \label{ell_full}
\end{equation}
\begin{equation}
2\frac{d\xi}{dt}=H\xi\left[2+\left(1+\frac{1}{\mu^{2}}\right)v^{2}\right]+v\left[k\left(1-\frac{1}{\mu^{2}}\right)+c\left(f_{0}+s\right)\right]
+\left[d_{m}\left(\ell\right)-1\right]\frac{\xi}{\ell}\frac{d\ell}{dt}\label{qui_full}
\end{equation}
\begin{equation}
\frac{dv}{dt}=\left(1-v^{2}\right)\left[\frac{k(v)}{\xi\mu^{2}}-Hv\left(1+\frac{1}{\mu^{2}}\right)-\frac{1}{1+\mu^{2}}\frac{\left[d_{m}\left(\ell\right)-1\right]}{v\ell}\frac{d\ell}{dt}\right]\label{v_full}
\end{equation}
\begin{equation}
\frac{1}{\mu}\frac{d\mu}{dt}=\frac{v}{\xi}\left[k\left(1-\frac{1}{\mu^{2}}\right)-c\left(f-f_{0}-s\right)\right]-H\left(1-\frac{1}{\mu^{2}}\right)
+\frac{\left[d_{m}\left(\ell\right)-1\right]}{\ell}\frac{d\ell}{dt}\label{mu_full}.
\end{equation}
\end{widetext}
We note that the Eqs. (\ref{ell_full}), (\ref{qui_full}) and (\ref{mu_full}) are related by Eq. (\ref{ximuL}), and therefore only two of them are independent. 

It is clear from these equations that the evolution of the network is driven by three main mechanisms: expansion, energy losses, and the choice of the scale in which wiggles are coarse-grained. Understanding the roles of all three is important to ascertain whether small-scale structure reaches scaling, i.e., whether $\mu$ evolves towards a constant value, and under what physical conditions this can happen. From a physical point of view, there are three possible scaling regimes: wiggliness can disappear (the trivial Nambu-Goto), reach scaling (becoming a constant) or grow. Numerical simulations \cite{FRAC} suggest that small-scale scaling is achieved at least in the matter-dominated era, with the interpretation in the radiation era being less clear. These questions provide the overall motivation for the exploration of the non-trivial solutions of the wiggly model.

As in Paper I, we carry out a systematic study of the asymptotic scaling solutions of the wiggly generalisation of the VOS. Specifically, we consider scaling solutions of the generic form
\be\begin{split}
    L &= \zeta_0 t^\alpha\\
    v &= v_0t^\beta\\
    \mu &= m_0t^\gamma\,;
\end{split}\ee
using Eq. (\ref{ximuL}), this also leads to
\be
\xi=\sqrt{m_0}\zeta_0 t^{\alpha+\gamma/2}\,.
\ee
For convenience, Table \ref{table1} summarizes the eleven different scaling solutions obtained in Paper I, and the main conditions under which they are valid.

\begin{table*}
  \begin{center}
    \begin{tabular}{|c|c|c|c|c|c|c|c|}
      \hline
Paper I & Expansion & Energy losses & $L$ & $v$ & $\mu$ & $\xi$ & Condition(s)\\
\hline
Eq. 37 & No & No & $L=\zeta_0$ & $v=v_0$ & $\mu=m_0$ & $\xi=\sqrt{m_0}\zeta_0$ & None\\
\hline
Eq. 38 & No & Yes & $ L=\zeta_0 t$ & $v=v_0$ & $\mu=m_0$ & $\xi=\sqrt{m_0}\zeta_0 t$ & $\eta\ge D$\\
Eq. 43 & No & Yes & $L= \zeta_0 t^{1-\gamma/2}$ & $v=v_0$ & $\mu=m_0t^\gamma$ & $\xi = \sqrt{m_0}\zeta_0 t$ & $\eta<D$, $\gamma=2\frac{D-\eta}{1+D}$\\
\hline
Eq. 49 & Yes & No & $L = \frac{k}{2\sqrt{\lambda(1-\lambda)}}t$ & $v = \sqrt{\lambda^{-1}-1}$ & $\mu =1$ & $\xi = L$ & $\lambda\ge\frac{2}{3}$\\
Eq. 50 & Yes & No & $L = \zeta_0t$ & $v = \frac{1}{\sqrt{1+m_0^2}}$ & $\mu = m_0$ & $\xi = \frac{3}{2}kv_0t$ & $\lambda=\frac{2}{3}$\\
Eq. 52 & Yes & No & $L = \zeta_0t^{3\lambda/2}$ & $v = v_0t^{-\lambda}$ & $\mu = m_0t^{2-5\lambda}$ & $\xi = \frac{kv_0}{2-4\lambda}t^{1-\lambda}$ & $\lambda\le\frac{1}{3}$\\
Eq. 55 & Yes & No & $L = \zeta_0t^{3\lambda/2}$ & $v = \frac{t^{\lambda-2/3}}{\sqrt{3\lambda-1}m_0}$ & $\mu = m_0t^{2/3-\lambda}$ & $\xi = \frac{3}{2}kv_0t^{\lambda+1/3}$ & $\frac{1}{3} <\lambda <\frac{2}{3}$\\
\hline
Eq. 63 & Yes & Yes & $L = \zeta_0t$ & $v = v_0$ & $\mu = 1$ & $\xi = L$ & $\lambda\ge\frac{2k}{3k+c}$\\
Eq. 66 & Yes & Yes & $L = \zeta_0t$ & $v = v_0$ & $\mu=m_0$ & $\xi = \frac{k}{\lambda v_0(1+m_0^2)}t$ & $\lambda=\frac{2k_{eff}}{3k_{eff}+c_{eff}}$\\
Eq. 81 & Yes & Yes & $L = \zeta_0t^{1-\lambda-\gamma/2}$ & $v = v_0t^{-\lambda}$ & $\mu = m_0t^{\gamma}$ & $\xi = \sqrt{m_0}\zeta_0t^{1-\lambda}$ &  $\lambda\le\frac{k_{eff}}{3k_{eff}+c_{eff}}$\\
Eq. 85 & Yes & Yes & $L = \zeta_0t^{1-3\gamma/2}$ & $v = v_0t^{-\gamma}$ & $\mu = m_0t^{\gamma}$ & $\xi = \sqrt{m_0}\zeta_0t^{1-\gamma}$ & $\frac{k_{eff}}{3k_{eff}+c_{eff}}<\lambda<\frac{2k_{eff}}{3k_{eff}+c_{eff}}$\\
\hline
    \end{tabular}
\caption{The eleven scaling solutions obtained in Paper I, together with the conditions under which they apply. The first column denotes the equation in Paper I in which the solution is first presented. The second and third columns indicate whether expansion and energy losses are assumed to be present. The next four columns denote (sometimes in a simplified way), the scaling laws for the VOS dynamical variables $L$, $v$, $\mu$ and $\xi$, defined in the text; quantities with an index 0 denote constants. The parameter $\lambda$ refers to the expansion rate of the universe, which is assumed to be $a(t)\propto t^\lambda$. The final column presents the main or simplest necessary conditions on the model parameters; these are necessary conditions for the solution to exist, but in some cases they are not sufficient (i.e., additional conditions involving the model parameters also apply).}
    \label{table1}
    \end{center}
\end{table*}

In what follows we extend the earlier work by considering a time-evolving coarse-graining scale. In particular, we consider a power-law shaped scale
\be
\ell = \ell_0 t^{\delta}\,.
\ee
More specifically, we consider strictly nonzero values of this power law, that is $0<\delta\le1$, with the upper limit being set by causality. Moreover, note that we can consider the case $\ell=\xi$ (corresponding to using the correlation length itself as the coarse-graining scale) by making the specific choice $\delta=\alpha+\gamma/2$. We also adopt the following phenomenological relation for the fractal dimension
\be
d_m(\mu)=2-\frac{1}{\mu^2}.
\ee

For convenience we also recall the definitions of the two effective parameters introduced in Paper I
\bq
k_{eff} &\equiv k +c(D-\eta) \label{defk}\\
c_{eff} &\equiv c(1+\eta).\label{defc}
\eq
These effective parameters are physically meaningful, as the existence of small-scale structure on the network modifies string curvature and further stimulates energy losses. Our analysis will follow the same structure as in Paper I, separately treating the solutions in Minkowski space and then for power law expanding universes. The reasoning for this separation is that the momentum parameter $k=k(v)$ is expected to vanish in the former case but is non-zero in the latter. Finally, we note that in our discussion of the scaling solutions we will generally denote the momentum parameter simply by $k$; its velocity dependence is not explicitly relevant since all the scaling solutions either imply $v=const.$ or $v\longrightarrow0$, and in both of these regimes $k(v)$ reduces to a constant value. Lastly, in what follows we discard ultra-relativistic $v=1$ solutions as these hold no physical relevance despite being mathematically allowed.

%%%%%%%%%%%%%%%%%%%%%%%%%%%%%%%%%%%%%%%%%%%%%%%%%%%%%%%%%%%%%%%%
\section{\label{flat}Scaling solutions without expansion}

Here we consider solutions in Minkowski space, by setting $H=0$. We recall that in this case we expect the VOS model to hold for a vanishing momentum parameter $k=0$ \cite{MS4,FRAC,Book,Correia1}.

\subsection{Without energy losses}

We start by considering the simplest case possible, with the only dynamical mechanism being the varying coarse-graining scale itself. It follows from Eq. (\ref{ell_full}) that the characteristic length scale of the network is constant, which is a direct consequence of the conservation of the total energy density of the network. We find two distinct scaling regimes, the first being the trivial Nambu-Goto solution
\be\label{case1}
\begin{split}
    L &= \zeta_0\\
    v &= v_0\\
    \mu &= 1\\
    \xi &= L\,;
\end{split}
\ee
this is analogous to the solution of Eq.(37) of Paper I, with the exception that the wiggliness is no longer arbitrary but restricted to the Nambu-Goto value $\mu=1$.
Not only do the scaling coefficients not exhibit any explicit dependence on the scale, but this solution also exists for any value of $\delta$. This solution trivially shows that in the absence of small-scale structure, a change of the coarse-graining scale has no impact on our description of the network.

The second solution is non-trivial, having growing small-scale structure
\be\label{case2}
\begin{split}
    L &= \zeta_0\\
    v &= v_0 t^{-\delta}\\
    \mu &= m_0 t^{\delta}\\
    \xi &= \sqrt{m_0} \zeta_0 t^{\delta/2}\,,
\end{split}
\ee
subject to the following constraint
\bq
v_0^2 m_0^2 =1. 
\label{case2a}
\eq
Equivalently this solution can be written, in terms of $\ell$, as
\be\label{case1b2}
v^{-1}\propto \mu\propto\ell\,, \xi\propto\sqrt{\ell}\,;
\ee
this solution, like the previous one, still exists for any value of $\delta \neq 0$. In the presence of small-scale structure, a growing coarse-graining scale leads to correspondingly larger wiggliness on that scale, which (as required by energy conservation, since no energy loss mechanisms are present) is compensated by a decreasing velocity on that same scale. This is one example of the point, already made in the previous section, in other words, that by varying the coarse-graining scale we are merely changing the way in which the network's energy is distributed between the bare string and the wiggles, with the total energy being conserved. 

Moreover, the condition imposed by Eq. (\ref{case2a}) suggests that, just like the wiggliness, the velocity is also a scale-dependent quantity. Therefore this also suggests that in the context of the wiggly VOS model, the velocity should be interpreted as a mesoscopic velocity rather than a microscopic RMS one. These two interpretations have been previously discussed in \cite{PAP1,PAP2}. 

We also note that the first solution can be obtained by taking the fixed scale limit $\delta \to 0$ of the second one. Finally, the specific case $\ell\propto\xi$ would correspond to $\delta=\gamma/2$, while the second solution has $\gamma=\delta$; together, these imply $\delta=\gamma=0$. One therefore concludes that for the choice of an averaging scale equal to the network's correlation length, $\ell\propto\xi$, there is no non-trivial scaling solution (other than the Nambu-Goto one).

%%%%%%%%%%%%%%%%%%%%%%%%%%%%%%%%%%%%%%%%%%%%%%%%%%%%%%%%%%%%
\subsection{With energy losses}

We now allow for the possibility of energy losses within the network. In particular, we assume the energy loss terms previously introduced in Eq. (\ref{enloss}). The presence of energy losses leads to a different dynamics of the characteristic length scale. We again find two possible solutions.

The first solution is simply the Nambu-Goto solution, implying linear scaling of both length scales
\be\label{case3}
\begin{split}
    L &= \frac{1}{2}cv_0t\\
    v &= v_0 \\
    \mu &= 1\\
    \xi &= L\,.
\end{split}
\ee
This solution is analogous to that of Eq.(38) of Paper I, again with the caveat that the constant $m_0$ is no longer arbitrary but restricted to the Nambu-Goto case, $\mu=1$. Clearly, the solution reflects the fact that if there is no small-scale structure on the strings, then a change in coarse-graining scale makes no difference.

The second scaling regime consists of a growing wiggliness solution, but now the characteristic lengthscale is also affected. The solution has the following form
\be\label{case4}
\begin{split}
    L &= \zeta_0 t^{\alpha}\\
    v &= v_0 t^{-\gamma} \\
    \mu &= m_0 t^{\gamma}\\
    \xi &=\sqrt{m_0}\zeta_0t^{1-\gamma}\,,
\end{split}
\ee
subject to the following conditions
\bq
\alpha&=&\frac{(\frac{2}{3}-\delta)(1+\eta)}{\frac{2}{3}(1+\eta)+2(D-\eta)} \label{alphaeq}\\ 
\gamma&=&\frac{2(D-\eta)+\delta(1+\eta)}{3(D-\eta)+(1+\eta)} \label{gammaeq}\\ 
\frac{cv_0}{\sqrt{m_0}\zeta_0}&=&\frac{2-3\delta}{3(D-\eta)+(1+\eta)}\\
\eta&<&D\\
0<\delta&=&\gamma v_0^2m_0^2<\frac{2}{3}\,. \label{deltaeq}
\eq
We note that in this second case taking the limit $\delta\to0$ effectively leads to solutions akin to that of Eq. (43) of Paper I, with $\gamma\to0$, $\alpha\to1$ and $(D-\eta)\to0$. A second novelty, as compared to the solutions described so far, is that $\delta$ is no longer arbitrary but limited by the condition $\delta<2/3$. This may be interpreted as a consequence of the energy losses of the network: if the coarse-graining scale does not evolve slowly enough, one is unable to find small-scale structure. For faster growing scales, the only mathematically allowed solution is the Nambu-Goto one, and no small-scale structure is seen. Physically, this means that all the energy is ascribed to the bare strings.

On the other hand, taking the limit of no energy losses $(D,\eta)\to0$ partially recovers the previous behavior of Eq. (\ref{case2}), and of Eq. (38) of Paper I. Specifically, the wiggliness and velocity have the expected complementary behaviour, with $-\beta=\gamma=\delta$, but the behaviour of $\alpha$ is now scale-dependent (with $\alpha=1-3\delta/2$, unlike Eq. (\ref{case2})), and consequently $\xi$ is no longer scaling linearly (instead we have $\xi\propto t^{1-\delta}$, unlike Eq. (38) of Paper I).

This explicit relation between $\delta$ and $\alpha$ in Eq. (\ref{alphaeq}) corresponds to an apparent scale dependence of the total energy of the network. As has already been mentioned, one would expect the total energy of the network to be a scale-invariant quantity, independent of the coarse-graining scale. (In other words, the choice of coarse-graining scale can affect how the energy is distributed between the bare string and the wiggles, but should not affect how the network loses energy.) This may simply indicate that this solution is unphysical for generic choices of $\delta$, or in other words, that scaling solutions of this kind do not exist. The only cases where such a solution is well-behaved are the limiting cases $\delta\to0$ and $\delta\to2/3$, which respectively lead to solutions with $\alpha=1$ and $\alpha=0$, both of which have been previously discussed.

Finally, if we specifically choose the scale to be that of the correlation length, $\ell\propto\xi$, we require $\delta=\alpha+\gamma/2=1-\gamma$ from which we find 
\bq
\delta &=& \frac{1 + D}{2(1+\eta)+3(D-\eta)}\\
\alpha&=&\frac{1}{2}\frac{(1+\eta)}{3(D-\eta)+2(1+\eta)}\\ 
\gamma&=&\frac{2(D-\eta)+(1+\eta)}{3(D-\eta)+2(1+\eta)}\\
\zeta_0\sqrt{m_0}&=& c v_0 \left[ 2(1+\eta)+3(D-\eta) \right]\\
v_0^2 m_0^2 &=& \frac{1+D}{ (1+\eta)+2(D-\eta)}\,.
\eq
Since we require that $\delta<2/3$, we are led to the following condition
\be
\eta<\frac{1}{2}(1+3D)\,
\ee
which is more stringent than the previous $\eta<D$, which also led to growing wiggliness solutions for $\delta=0$, cf. Eq. (38) in Paper I. The physical interpretation of this stronger bound is clear; the small-scale structure losses must not only be smaller than the gains, but must be smaller enough such that the small-structure can still grow if the coarse-graining scale is also growing.

We can additionally take the limit of no energy loss, $(D-\eta)\to0$, which yields
\bq 
L\propto\sqrt{\ell}\propto t^{1/4}\,, v^{-1}\propto\mu\propto\xi\propto\ell\propto t^{1/2}\,.
\label{case4a}
\eq
Unlike the case of no energy losses, here there is a possible solution with a time-evolving coarse-graining scale, but this is not a linear scaling solution. Thus, in this case there is still no linear scaling other than the trivial Nambu-Goto one. 

%%%%%%%%%%%%%%%%%%%%%%%%%%%%%%%%%%%%%%%%%%%%%%%%%%%%%%%%%%%%%%%%%%%%%%%
\section{\label{expand}Scaling solutions in expanding universes}

We now consider solutions in expanding universes, and more precisely, power-law expanding universes of the form $a(t)\propto t^\lambda$, with $0<\lambda<1$. In this case, one expects $k\neq0$ as confirmed by numerical simulations \cite{FRAC}. Following the structure of the previous section, we first examine the case without energy losses and then the more realistic case where they are allowed.

\subsection{Without energy losses}

In this case, the three classes of scaling regimes are possible. For the sake of clarity, we describe each in a separate  subsubsection.

\subsubsection{Nambu-Goto solution}

Firstly, we have the Nambu-Goto scaling regime corresponding to Eq. (49) of Paper I
\be\label{case5}
\begin{split}
    L &= \frac{k}{2\sqrt{\lambda(1-\lambda)}}t\\
    v &= \sqrt{\lambda^{-1}-1}\\
    \mu &=1\\
    \xi &= \frac{k}{2\sqrt{\lambda(1-\lambda)}}t\,,
\end{split}
\ee
with the same restrictions as in the $\delta=0$ case. If we interpret this velocity as a microscopic one and only require $v_0^2 < 1$, this condition restricts the expansion rates to $ \lambda > 1/2$, while if we interpret it as an average or mesoscopic one, the mean loop velocity in Minkowski space leads to the requirement that $v_0^2 \leq 1/2$, which makes this solution physically viable only in the range $\lambda \geq 2/3$. In either case, slower expansion rates do not ensure sufficient Hubble damping for the network to reach linear scaling. It follows that scaling can not be reached in the radiation-dominated era, as the network simply does not lose sufficient energy, but can be attained in the matter era. Again the interpretation of this solution is straightforward: since there is no small-scale structure, the introduction of a time-dependent averaging scale neither affects the scaling behavior of the VOS solution, nor does it impose any restriction on $\delta$.

\subsubsection{Full scaling}

Secondly, there is also a constant wiggliness solution
\be\label{case6}
\begin{split}
    L &= \zeta_0 t\\
    v &= v_0 \\
    \mu &=m_0\\
    \xi &= \frac{kv_0}{\lambda+\delta}t\,.
\end{split}
\ee
subject to the following constraints
\bq
v_0^2&=&\frac{1+(2\lambda^{-1}-3)m_0^2}{1+m_0^2}\\
\delta&=&\left(1-\frac{3}{2}\lambda\right)m_0^2(1+m_0^2) \label{limits3_delta}\\
max&&\left(\frac{1}{2},\frac{2}{3}\left[1-\frac{1}{m_0^2(1+m_0^2)}\right]\right)<\lambda<\frac{2}{3}\,.\label{limits3}
\eq

\begin{figure}
\centering
  \includegraphics[width=1.00\columnwidth]{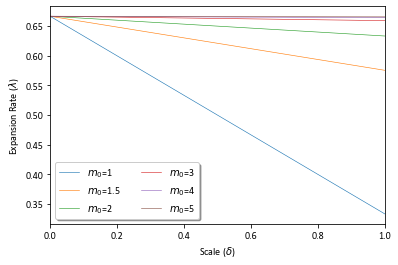}
  \caption{The expansion rate $\lambda$ as dictated by Eq. (\ref{limits3_delta}), for various values of the constant wiggliness $m_0$, as a function of the averaging scale exponent $\delta$. }
  \label{fig1}
\end{figure}

Note that Eq. (\ref{case6}) shows that the correlation length, whose normalization is inversely proportional to $(\lambda+\delta)$, becomes smaller for faster growing coarse-graining scales. For $\delta=0$ one has $\lambda=2/3$ and therefore $\xi=(3/2)kv_0t$. This is clearly an extension of Eq. (50) of Paper I, to which it exactly reduces if we take the limit $\delta\to0$. It has been shown in Paper I that a full scaling regime of this kind, with no other dynamical mechanisms acting on the network, was only possible in the matter era. It is physically fully  consistent that the inclusion of a growing coarse-graining scale decreases the value of the expansion rate for which this kind of regime can occur. This is illustrated in Fig. \ref{fig1}, along with the increase of the allowed expansion rate with the wiggliness.

\begin{figure}
\centering
  \includegraphics[width=1.00\columnwidth]{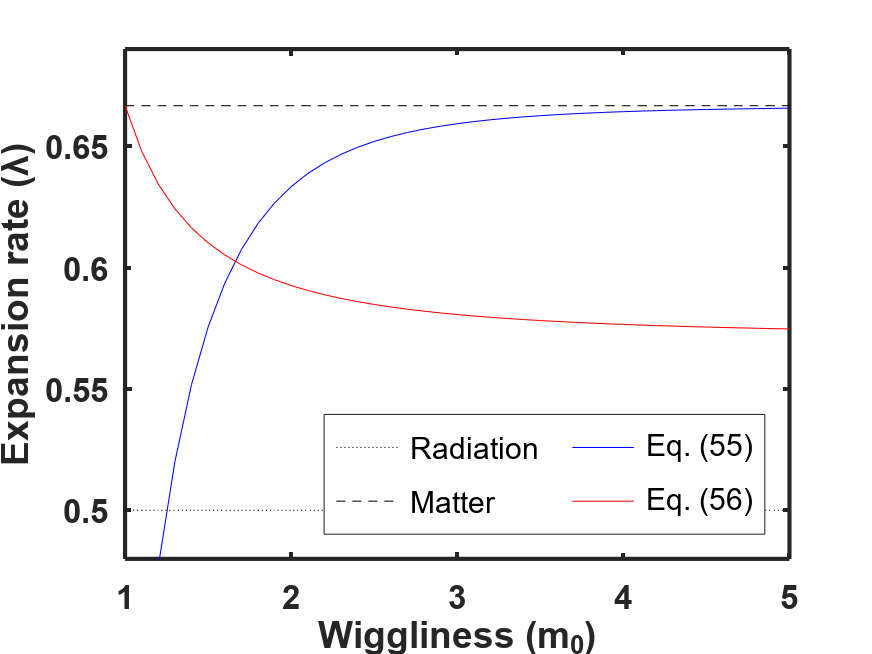}
  \caption{The range of expansion rates $\lambda$ allowed by Eq. (\ref{limits3}) and Eq. (\ref{limits47}), as a function of the constant wiggliness $m_0$, respectively in blue and red lines. In each case the allowed range is between the black dotted and dashed lines (which denote the radiation and matter eras respectively), and above the corresponding colored line.}
  \label{fig2}
\end{figure}

In the last constraint, the upper limit comes from requiring $\delta>0$, while the lower limits come from requiring $v_0^2<1$ and $\delta\le1$ (and therefore the first of the upper limits is strict while for the second, the equality is allowed). In the low wiggliness limit ($m_0\to1$) all the range of expansion rates between radiation and matter era are allowed, while in the large wiggliness limit ($m_0\to\infty$) no expansion rate is allowed, cf. Fig. \ref{fig2}.

%%%%%%%%%%%%%%%%%%%%%%%%%%%%%%%%%%%%%%%%%%%%%%%
%%%%%%%%%%%%%%%%%%%%%%%%%%%%%%%%%%%%%%%%%%%%%%%

The last condition, Eq, (\ref{limits3}), implies that this solution can exist, at most, for expansion rates between the radiation and matter eras, $1/2<\lambda<2/3$. However, note that this relies on the velocity limit being assumed to be $v_0^2<1$. If instead one requires $v_0^2<1/2$, then one gets the condition
\be
\lambda>\frac{4m_0^2}{7m_0^2-1}\,; \label{limits47}
\ee
here the allowed parameter space is to some extent complementary to that of the last condition in Eq. (\ref{limits3}), cf. Fig. \ref{fig2}. In the limit $m_0\to1$ this reduces to $\lambda>2/3$, implying that with this assumption on the velocity, this time-dependent coarse-graining scale extension of Eq. (50) of Paper I is not physically allowed, and therefore that there is no full scaling solution for the network (other than the Nambu-Goto one, which has $\mu=1$). On the other hand, as the value of $m_0$ increases, so does the range of allowed expansion rates.

%%%%%%%%%%%%%%%%%%%%%%%%%%%%%%%%%%%%%%%%%%%%%%%
%%%%%%%%%%%%%%%%%%%%%%%%%%%%%%%%%%%%%%%%%%%%%%%

If we assume that Eq. (\ref{limits3}) holds and therefore that this generalisation of Eq. (50) of Paper I does exist, we can specifically choose the scale to be that of the correlation length. In that case we must have $\delta=1$, which leads to the condition
\be
m_0^2(1+m_0^2)=\frac{2}{2-3\lambda}\,.
\ee
Specifically, for the radiation era ($\lambda=1/2$) one finds
\be
m_0=\sqrt{\frac{1}{2}(\sqrt{17}-1)}\sim1.25\,,
\ee
which is commensurate with the results of radiation era numerical simulations \cite{FRAC}, although we emphasize the above solution does not include energy losses.

\subsubsection{Growing wiggliness}

Lastly, there are solutions where small-scale structure is allowed to grow, which have exactly the same requirements as the expansion-only case, that is $\alpha=3\lambda/2$ (thereby imposing the physical constraint $\lambda < \frac{2}{3}$), $\beta < 0$ (decaying velocities), together with $3\lambda/3 -\beta+\gamma/2=1$, and $\beta+\gamma \geq 0$. These are extensions to the solutions given by Eqs. (52) and (55) of Paper I, for slow and intermediate expansion rates, which indeed are only minimally changed.

For slow expansion rates we have
\be\label{case7}
\begin{split}
    L &= \zeta_0 t^{3\lambda/2}\\
    v &= v_0 t^{-\lambda}\\
    \mu &=m_0 t^{2-5\lambda}\\
    \xi &= \frac{kv_0}{2-4\lambda+\delta}t^{1-\lambda}\,.
\end{split}
\ee
with the conditions
\bq
    \sqrt{m_0 }\zeta_0 &=& \frac{k v_0}{2-4\lambda+\delta}\\
    \lambda &\leq& \frac{1}{3}\,,
\eq
which in the limit $\delta \to0$ trivially recovers Eq. (52) of Paper I.

In the intermediate expansion rate regime, which includes the radiation-dominated era but not the matter era, one has
\be\label{case8}
\begin{split}
    L &= \zeta_0 t^{3\lambda/2}\\
    v &= \frac{t^{\lambda-2/3}}{\sqrt{3\lambda-1}m_0}\\
    \mu &=m_0 t^{2/3-\lambda}\\
    \xi &= \frac{kv_0}{\delta+2/3}t^{\lambda+1/3}\,.
\end{split}
\ee
subject to the constraints
\bq
\sqrt{m_0}\zeta_0&=&\frac{kv_0}{\delta+2/3}\\
\frac{1}{3}<&\lambda&<\frac{2}{3}\,.
\eq
Once again, this solution is a straightforward  generalization of Eq. (55) of Paper I, to which it reduces to in the limit $\delta \to0$. We also note that the slow and intermediate scaling regimes match for an expansion rate of $\lambda =1/3$.

Note that in both branches of these growing wiggliness solution, the main impact of the increasing coarse-graining scale is that the correlation length $\xi$ becomes smaller, since $\delta$ appears in the denominator of its normalization. As previously mentioned, this is also the case for the solution given by Eq. (\ref{case6}): the faster the coarse-graining scale grows, the smaller the correlation length.

If in this growing wiggliness case we choose the scale to be that of the correlation length, $\ell\propto\xi$, we have the following simplification in the slow expansion rates
\bq
\delta&=&1-\lambda\\
\sqrt{m_0}\zeta_0&=&\frac{kv_0}{3-5\lambda}\\
\lambda&\le&\frac{1}{3}\,,
\eq
while the corresponding relations for the intermediate expansion rates are
\bq
\delta&=&\frac{1}{3}+\lambda\\
\sqrt{m_0}\zeta_0&=&\frac{kv_0}{1+\lambda}\\
\frac{1}{3}<&\lambda&<\frac{2}{3}\,;
\eq
as expected the two solutions match for $\lambda=1/3$. Overall, this implies that for these solutions $\delta$ has a minimum value of $\delta=2/3$ (for the transition $\lambda=1/3$ case) and approaches $\delta=1$ in the limits $\lambda\to0$ and $\lambda\to2/3$. In both of these cases there is only one choice of coarse-graining scale $\delta$, for each expansion rate $\lambda$, that leads to this solution.

\subsection{With energy losses}

Finally, we consider the most realistic case which includes all the three dynamical mechanisms: expansion, energy losses and a varying coarse-graining scale. In this case, the three scaling regimes considered in the previous subsection can in principle exist, and are extensions of them.

\subsubsection{Nambu-Goto solution}

Firstly, we have the canonical VOS Nambu-Goto solution
\be\label{case9}
\begin{split}
    L &= \zeta_0t\\
    v &= v_0\\
    \mu &= 1\\
    \xi &= \zeta_0t\,,
\end{split}
\ee
with the scaling parameters being given by
\bq
\zeta^2_{NG}&=&\frac{k(k+c)}{4\lambda(1-\lambda)}\\
v^2_{NG}&=&\frac{(1-\lambda)k}{\lambda(k+c)}\,.
\eq
Again, in this case, since the network contains no wiggliness, a growing coarse-graining scale makes no difference; this solution is therefore the same of Eq. (63) of Paper I. In the $c=0$ limit we recover the previous solution given by Eq. (\ref{case5}). The consistency condition relating the expansion rate and the VOS model parameters is $\lambda>k/(2k+c)$ if one only requires $v_0^2<1$, or $\lambda\ge 2k/(3k+c)$ if one imposes $v_0^2\le1/2$.

\subsubsection{Full scaling}

There is also a full linear scaling solution
\be\label{case10}
\begin{split}
    L &= \zeta_0 t\\
    v &= v_0\\
    \mu &= m_0\\
    \xi &= \sqrt{m_0} \zeta_0 t\,,
\end{split}
\ee
subject to the following consistency relations:
\bw
\be
\footnotesize
\zeta_0^2 =  \frac{\left( c \eta m_0^{3/2}-c(1+\eta)m_0^2 \right) \left( \delta (1-m_0^{-2}) ( c(1+\eta)m_0^4 - c\eta m_0^{7/2} + k m_0^2) - k(1+m_0^2)(\lambda+(2-3 \lambda) m_0^2)  \right) + k^2(m_0^2+1)\left(\lambda+(2-3 \lambda) m_0^2 \right) }{ \lambda m_0 \bigg( \delta(m_0^2-1) + (m_0^2+1)(\lambda+(2-3 \lambda) m_0^2) \bigg)^2}
\ee
\be
v_0^2 = \frac{\delta (1-m_0^{-2}) \left( c \eta m_0^{7/2} - c(1+\eta)m_0^4 \right) + k(m_0^2+1) \left(\lambda+(2-3 \lambda) m_0^2 \right) }{\lambda (1+ m_0^2) \left[ k(m_0^2+1) + (m_0^{-2}+1) ( c(1+\eta)m_0^4 - c\eta m_0^{7/2} )   \right]}
\ee
\be
\begin{split}
\bigg( \lambda+m_0^{2}(2-3 \lambda)\bigg) (1+ m_0^2) \bigg( (k+c D)(1-\frac{1}{m_0^{2}})  -c \eta (1-\frac{1}{m_0^{1/2}} ) \bigg) =\lambda \bigg(1-\frac{1}{m_0^{2}} \bigg) \bigg( k(m_0^2+1)+ \\
+ (m_0^{-2}+1) ( c(1+\eta)m_0^4 - c\eta m_0^{7/2} ) \bigg)
-  \delta \left(1- \frac{1}{m_0^2} \right) \bigg( c(1+\eta) m_0^4 + (2k + c+cD)m_0^2 - c\eta m_0^{7/2} -cD \bigg)\,.
\end{split}
\ee
\ew
We note this solution is a generalization of Eq. (66) of Paper I, to which it reduces to in the limit $\delta \to0$. Moreover, setting $m_0=1$ the first two conditions recover the canonical Nambu-Goto solution, Eq. (\ref{case9}), while the third condition becomes trivial.

%%%%%%%%%%%%%%%%%%%%%%%%%%%%%%%%%%%%%%%%%%%%%%%
\begin{figure}
\centering
  \includegraphics[width=1.00\columnwidth]{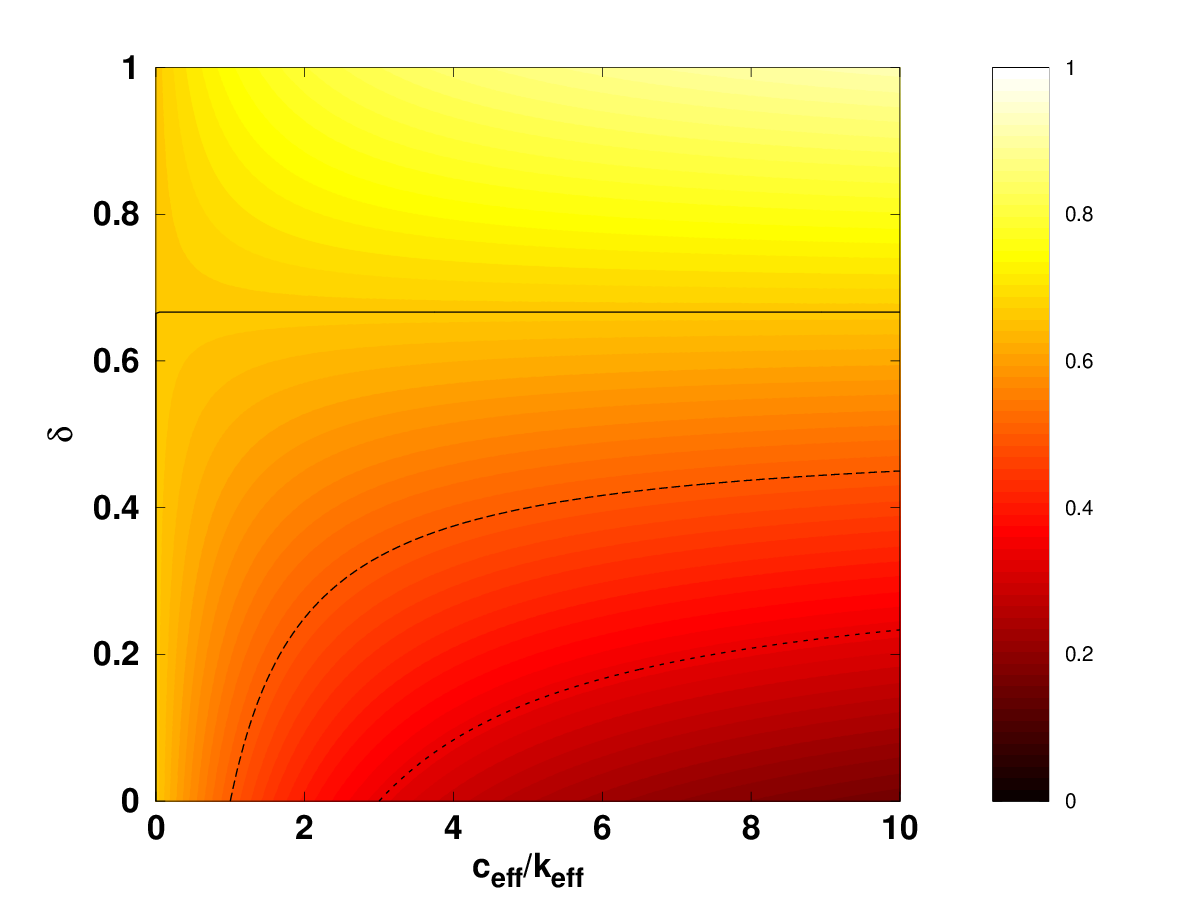}
  \caption{The expansion rate $\lambda$ given by Eq. (\ref{condition23d}), given by the colormap, as a function of the ratio of the phenomenological parameters $c_{eff}/k_{eff}$ and the exponent of the time-dependent coarse-graining scale $\delta$. For convenience the particular values $\lambda=1/3$, $\lambda=1/2$ (radiation era) and $\lambda=2/3$ (matter era) are shown with dotted, dashed and solid black lines respectively.}
  \label{fig3}
\end{figure}
%%%%%%%%%%%%%%%%%%%%%%%%%%%%%%%%%%%%%%%%%%%%%%%

Despite the algebraic complexity of these conditions, it is straightforward to verify that, in the limit of large wiggliness, $m_0\to\infty$, this solution only exists for a single expansion rate
\be
\lambda=\frac{2 k_{eff} + \delta c_{eff}}{3 k_{eff}+c_{eff}}\,;\label{condition23d}
\ee
the corresponding scaling solution can be written
\be
\begin{split}
    L &= \sqrt{\frac{c_{eff}[(2-3\lambda)k-\delta c_{eff}]}{\lambda(2-3\lambda)^2}}\frac{t}{m_0^{3/2}}\\
    v &= \sqrt{\frac{(2-3\lambda)k-\delta c_{eff}}{\lambda c_{eff}}}\frac{1}{m_0}\\
    \mu &= m_0\\
    \xi &= \sqrt{\frac{c_{eff}[(2-3\lambda)k-\delta c_{eff}]}{\lambda(2-3\lambda)^2}}\frac{t}{m_0}\,.
\end{split}
\ee
Note that Eq. (\ref{condition23d}) implies that in order to have this scaling solution at the radiation-domination epoch we require
\be
k_{eff}=(1-2\delta)c_{eff}\,.
\ee
Both of these generalize the result of Paper I, which is recovered when $\delta\to0$. Note that, given physically reasonable values of the other parameters, such a scaling solution in the radiation era is only possible for $\delta<1/2$. On the other hand, requiring that this solution occurs in the matter era leads to $\delta=2/3$. In particular, this means that in this large wiggliness limit this solution is not possible for a coarse-graining scale $\ell\propto\xi$, since that would require $\delta=1$.

From  Eq. (\ref{condition23d}) one also sees that in the limit $c_{eff}/k_{eff}\to0$ one recovers $\lambda=2/3$ (the matter era, as expected), while in the opposite limit of $c_{eff}/k_{eff}\to\infty$ one would have $\lambda=\delta$. Figure \ref{fig3} illustrates all the above behaviours.

\subsubsection{Growing wiggliness}

Last but not least, the regime with growing wiggliness again contains two solutions depending on the expansion rate. These are analogous to the ones in Eqs. (\ref{case7}) and (\ref{case8}), and still require $\beta<0$ (decaying velocities), $\beta+\gamma\ge0$, and $\alpha-\beta+\gamma/2=1$, although $\alpha= 3\lambda/2$ no longer holds. Instead we require $\alpha> 3\lambda/2$, which is expected from a physical point of view since the inclusion of energy losses implies that the total energy of the network will decay faster. Moreover, both $\alpha$ and $\lambda$ depend on the VOS model parameters, and there are general consistency relations which apply to both solutions,
\be
\frac{v_0}{\sqrt{m_0}\zeta_0}=\frac{\lambda+\gamma-\delta}{k_{eff}}=\frac{2\alpha-3\lambda}{c_{eff}}\,,
\ee
with $k_{eff}$ and $c_{eff}$ as defined in Eq. (\ref{defk}) and Eq. (\ref{defc}) respectively. Note that in addition to the explicit presence of $\delta$ in these consistency relations, the scaling exponent $\gamma$ and (possibly) $\alpha$ also depend on $\delta$, so the impact of an increasing coarse-graining scale is not immediately clear from these relations.

In the slow expansion regime we have
\be\label{case11}
\begin{split}
    L &= \zeta_0 t^{\alpha} \\
    v &= v_0 t^{-\lambda} \\
    \mu &= m_0 t^{\gamma}\\
    \xi &= \sqrt{m_0} \zeta_0 t^{1-\lambda} \,
\end{split}
\ee
together with
\bq
\alpha&=&\frac{1}{2} \frac{(2- \delta) c_{eff} +\lambda(3k_{eff}-c_{eff})}{k_{eff}+c_{eff}}\\
\gamma&=& \frac{\delta c_{eff} + 2k_{eff}-\lambda(c_{eff}+5k_{eff} )}{k_{eff}+c_{eff}}\\ 
\lambda &\leq& \frac{1}{2}\frac{2k_{eff}+ \delta c_{eff}}{3 k_{eff} + c_{eff}} \label{lambda_ineq}\,,
\eq
which simplifies to Eq. (81) of Paper I in the fixed scale limit $\delta \to0$. Eliminating energy losses by taking the limit $(D-\eta) \to0$ also yields Eq. (\ref{case7}). Note the evolution of both the total energy of the network and the bare string energy is unchanged by the introduction of the averaging scale, being given by
\be
\frac{\rho_0}{\rho_{crit}} \propto t^{2 \lambda}
\ee
\be
\frac{\rho}{\rho_{crit}} \propto t^{2\lambda + \gamma}.
\ee

For intermediate expansion rates we have
\be\label{case12}
\begin{split}
    L &= \zeta_0 t^{\alpha} \\
    v &= v_0 t^{- \gamma} \\
    \mu &= m_0 t^{\gamma}\\
    \xi &= \sqrt{m_0} \zeta_0 t^{1-\gamma} \,
\end{split}
\ee
together with
\bq
\alpha&=& \frac{(\frac{2}{3}-\delta)c_{eff} + \lambda ( 3k_{eff} +c_{eff})}{\frac{2}{3}c_{eff}+2k_{eff}}\label{alpha_interm}\\
\gamma&=&\frac{ c_{eff} \delta + 2 k_{eff} - \lambda ( 3k_{eff} +c_{eff})}{c_{eff}+3 k_{eff}}\label{gamma_interm} \\ 
\lambda &\in& \bigg] \frac{1}{2}\frac{2k_{eff}+ \delta c_{eff}}{3 k_{eff} + c_{eff}}, \frac{2 k_{eff} + \delta c_{eff}}{3 k_{eff}+c_{eff}} \bigg[ \label{lambda_interm}\\
\delta &=& (2\alpha-3\lambda)\frac{k}{c_{eff}} + \frac{2}{3}\left(1-\alpha-\frac{3}{2}\lambda\right)v_0^2 m_0^2 \label{delta_interm}\,,
\eq

Again, we note that in the fixed scale limit $\delta \to0$, we recover Eq. (85) of Paper I. Further, in the no energy losses limit we are able to recover the previous solution Eq. (\ref{case8}). This solution implies that the ratio of bare string energy in relation to the background energy density evolves according to
\be
\frac{\rho_0}{\rho_{crit}} \propto t^{2 \gamma}
\ee
while for the total energy density of the network we have
\be
\frac{\rho}{\rho_{crit}} \propto t^{3\gamma}\,;
\ee
again, these are the same as for the analogous $\delta=0$ cases of these solutions, discussed in Paper I.

\begin{figure}
    \centering
    \includegraphics[width=1.00\columnwidth]{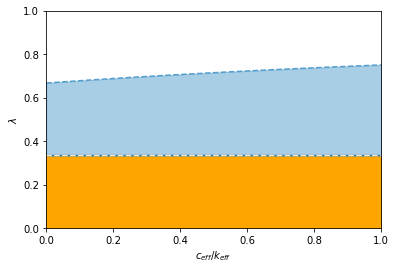}
    \caption{Range of allowed expansion rates as a function of $\frac{c_{eff}}{k_{eff}}$ in the case where $\ell \propto \xi$, in the slow (orange) and intermediate (blue) growing wiggliness regimes, according to Eqs. (\ref{lambda_ineq}) and (\ref{lambda_interm}) respectively.}
    \label{fig4}
\end{figure}

\begin{figure}
    \centering
    \includegraphics[width=1.00\columnwidth]{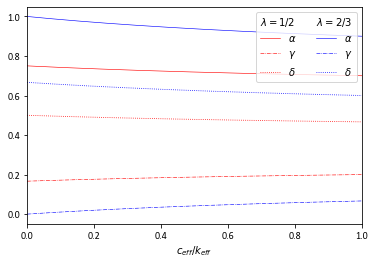}
    \caption{The values of the scaling exponents of the characteristic lengthscale (solid line), the wiggliness (dashed line) and the coarse-graining scale (dotted line) as a function of $\frac{c_{eff}}{k_{eff}}$, in the intermediate expansion rate regime, given by Eqs. (\ref{alpha_interm}), (\ref{gamma_interm}) and (\ref{delta_interm}) respectively, for the case where $\ell \propto \xi$. Red lines correspond to the matter era, while blue lines represent the radiation era.}
    \label{fig5}
\end{figure}

As before, we consider the specific case where $\ell \propto \xi$ for both of these solutions. In the slow expansion regime, it is clear that this choice of scale limits the expansions for which this type of scaling can occur, excluding both the matter and radiation eras. However, as seen in Fig. \ref{fig4}, the scaling solution described by Eq. (\ref{case12}) is expected to occur for intermediate expansion values, which include the matter and radiation eras. The behavior of the network in these two cosmological epochs is depicted in Fig. \ref{fig5}. This illustrates how full scaling of the network can take place in the matter era in the absence of energy losses, while this never occurs in the radiation era regardless of the values of the effective energy loss parameters.

%%%%%%%%%%%%%%%%%%%%%%%%%%%%%%%%%%%%%%%%%
%%%%%%%%%%%%%%%%%%%%%%%%%%%%%%%%%%%%%%%%%
%%%%%%%%%%%%%%%%%%%%%%%%%%%%%%%%%%%%%%%%%

A possible physical caveat to these mathematically allowed solutions is again the dependence of the scaling exponent $\alpha$ on the coarse-graining scale power $\delta$. In order to probe the isolated effect of the averaging scale on the evolution of the network, we can asymptotically eliminate the energy losses and Hubble damping by taking mathematical limits of the slow expansion regime solution, Eq. (\ref{case11}). If we do this by simultaneously assuming $\lambda\to0$, $k\to0$ and $(D-\eta)\to0$, we find $\alpha=1-\delta/2$ and $\gamma=\delta$, which corresponds to $\xi\propto t$, but this is again physically problematic due the apparent violation of energy conservation. On the other hand, if we assume $\lambda\to0$ and $c\to0$, we find the physically more acceptable $\alpha=0$ (expected since there are no explicit energy loss mechanisms), $\gamma=2$, and also $\xi\propto t$, independently of $\delta$.

This behaviour should be compared to the incoherent behavior previously found when considering the no energy losses limit in Eq. (\ref{case4}): taking $(D-\eta) \longrightarrow 0$ in Eq. (\ref{case4}) resulted in a solution partially similar but not identical to Eq. (\ref{case2}). That solution would correspond to a network losing energy, in spite of the absence of mechanisms that could explain this. The reason behind this discrepancy remains unclear, but it seems to be related to the role of the momentum parameter. The fact that in Minkowski space one sets the momentum parameter as $k=0$, seems to prevent $\alpha$ from reaching $0$ whenever the solution approaches the limit of no energy losses, deeming the solutions physically irrelevant.

\section{\label{concl}Conclusions}

\begin{figure*}
    \centering
    \includegraphics[width=1.0\textwidth]{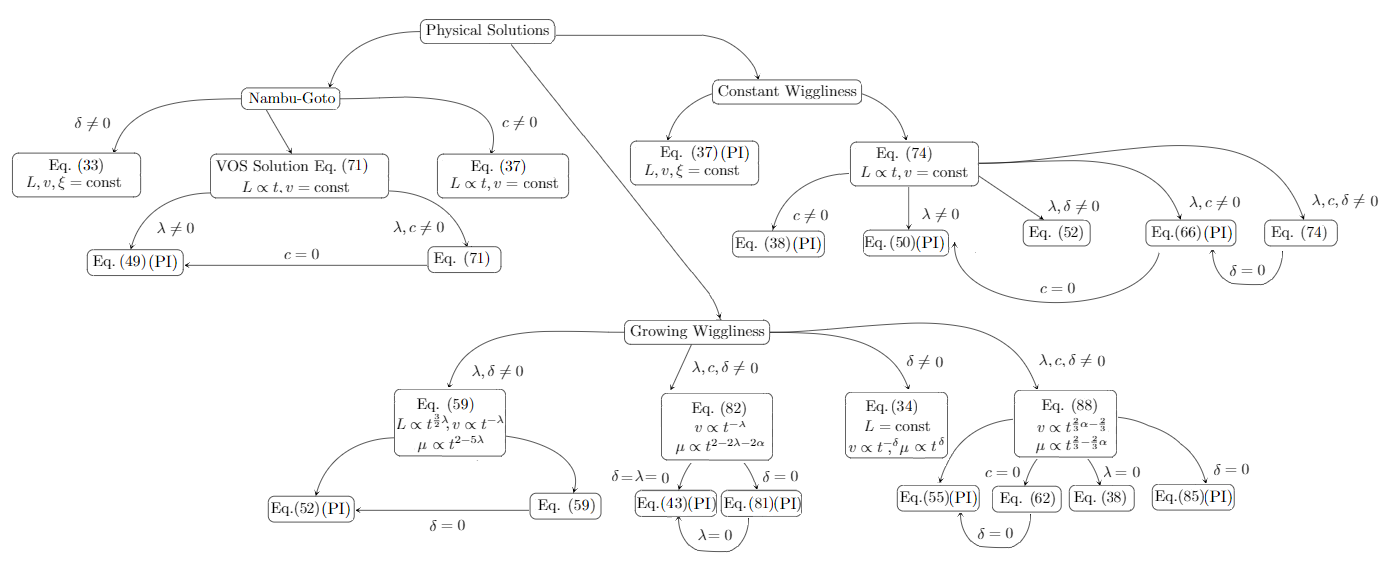}
    \caption{Schematic representation of the various families of scaling solutions. Here, PI stands for the solutions first described in in Paper I \cite{almeida}.}
    \label{fig6}
\end{figure*}

We have refined the physical interpretation of the wiggly generalisation of the VOS model \cite{PAP1,PAP2} by determining the possible scaling regimes and their allowed physical ranges and applicable consistency conditions. In doing this, we have also  further clarified the role of the various physical mechanisms on the evolution of the network. This analysis has been particularly focused on the role of the coarse-graining scale, since the other relevant physical mechanisms were already discussed in Paper I \cite{almeida}. For convenience, the landscape of all the possible scaling solutions of the wiggly model, including both the ones presented in Paper I (and summarized in Table \ref{table1} of the present work) and the additional ones which have been presented in the previous sections, is presented in schematic form in Figure \ref{fig6}.

%%%%%%%%%%%%%%%%%%%%%%%%%%%%%%%%%%%%%%%%%%%%%
%%%%%%%%%%%%%%%%%%%%%%%%%%%%%%%%%%%%%%%%%%%%%

In the absence of energy loss mechanisms, one expects the network to be in a trivial equilibrium solution. The inclusion of a growing time-dependent coarse-graining scale leads to a new scaling regime where small-scale structure grows in time, but this is compensated by a decreasing velocity, with the total energy density remaining constant. This behavior is consistent with the interpretation of a mesoscopic velocity that is embodied in the wiggly extension of the VOS model, rather than the traditional microscopic RMS velocity of the simplest version of the model. Further, this scaling solution allows the averaging scale to remain unbounded, which can be attributed to total energy conservation. If, however, the network is also subject to energy losses, the time dependence of the coarse-graining scale becomes bounded from above. The fact that its exponent is not allowed to take too large values should be interpreted as a consequence of the network losing energy: choosing a fast-growing coarse-graining scale would imply we are unable to find small-scale structure. Furthermore, while linear scaling in Minkowski space had been found in Paper I, in agreement with numerical simulations \cite{Sakellariadou,FRAC}, it is clear that the introduction of a time-varying coarse-graining scale prevents this type of scaling from occurring.

In power-law expanding universes, one expects the three scaling regimes for the wiggliness to be possible, depending on the expansion rate. For fast expansion rates, the Nambu-Goto solution subsists as wiggliness is unable to accumulate in the network. Slower expansion rates allow for either growing or constant wiggliness. In the latter case, which can occur for a single expansion rate (which depends on the remaining model parameters), the network is also expected to reach full scaling, where the energy density, velocity and wiggliness evolve towards a constant. In the absence of other energy loss mechanisms, this takes place in the matter-dominated era, as seen in Paper I. Allowing the coarse-graining scale to grow (with other phenomenological parameters unchanged) decreases the value of the expansion rate for which full scaling occurs. In this more general parameter space, it is no longer necessarily the case that full scaling is more likely to occur in the matter-dominated era than for other expansion rates. In particular, there are choices of phenomenological model parameters for which full scaling would occur in the radiation-dominated era. We note that the existence of three broad classes of scaling solutions is a generic feature of models of cosmic string networks with additional degrees of freedom on the string worldsheet, having previously been identified in chiral superconducting strings \cite{Oliveira} and more recently in general models possessing arbitrary currents and charges \cite{CVOS}.

Finally, in our analysis in the present work it has become clear that there is a new class of solutions that depict unexpected behavior, either due to  the lack of energy conservation whenever there are no mechanisms that could make the network lose energy (e.g., no Hubble damping, no production of loops or kinks), or the apparent discrepancies that arise when considering limit cases. We suggest that these solutions (e.g., Eq. \ref{case4}) should be treated as physically irrelevant, although they are mathematically allowed. The fact that such solutions only emerge whenever the coarse-graining scale is varying in time suggests that the interplay between coarse-graining and energy loss mechanisms (some of which are phenomenologically added to the equations, as opposed to being derived \textit{ab initio} from the relevant microscopic equations of motion) warrants additional study.

We conclude on a point already made in Paper I: our mapping of the landscape of the allowed scaling solutions of wiggly string networks paves the way for the detailed testing of the model, to be done by forthcoming high-resolution field theory and Nambu-Goto simulations. Some data from an earlier generation of Nambu-Goto simulations already exists \cite{BB,AS,Sakellariadou,FRAC,Blanco}, and while our results are in qualitative agreement with these works, their data is far from being sufficiently precise to allow a meaningful and quantitative comparison, not least because in most such simulations no wiggliness measurements are reported---the exception being \cite{FRAC}.

Our work therefore provides motivation for additional, higher resolution simulations. Nambu-Goto simulations will be particularly suitable for exploring the landscape of scaling solutions, since in these simulations one can switch intercommuting and loop production on and off at will. Nevertheless, one can envisage a similar analysis being done with Abelian-Higgs (field theory) simulations. While this would not be possible with traditional CPU-based simulations \cite{ABELIAN,Stuckey,Hiramatsu} due to a lack the spatial resolution and dynamic range to study small-scale wiggliness, a new generation GPU-accelerated Abelian-Higgs simulation code \cite{GPU1,GPU2} has emerged, enabling a detailed and statistically more robust calibration of the VOS model \cite{Correia1,Correia2,Correia3} and making a quantitative characterization of scales significantly below the correlation length possible. Such an analysis will also provide a cross-check of Nambu-Goto simulation results. Work along these lines is ongoing in our team.

%%%%%%%%%%%%%%%%%%%%%%%%%%%%%%%%%%%%%%%%%%%%%

\begin{acknowledgments}
This work was financed by FEDER---Fundo Europeu de Desenvolvimento Regional funds through the COMPETE 2020---Operational Programme for Competitiveness and Internationalisation (POCI), and by Portuguese funds through FCT - Funda\c c\~ao para a Ci\^encia e a Tecnologia in the framework of the project POCI-01-0145-FEDER-028987 and PTDC/FIS-AST/28987/2017.  CJM also acknowledges FCT and POCH/FSE (EC) support through Investigador FCT Contract 2021.01214.CEECIND/CP1658/CT0001. 
\end{acknowledgments}

\bibliography{ana}
\end{document}